\documentstyle[12pt,aaspp4,epsf]{article}

\def\etal{{\it et al.}} \def\ie{{\it i.e.}} \def\eg{{\it e.g.}} 
\def\lap{\hbox{${_{\displaystyle<}\atop^{\displaystyle\sim}}$}} 
\def\gap{\hbox{${_{\displaystyle>}\atop^{\displaystyle\sim}}$}}

\begin{document}
	
\title{Superfluid Friction and Late-time Thermal Evolution of Neutron 
Stars} \author{Michelle B. Larson} \affil{mlarson@physics.montana.edu} 
\and \author{Bennett Link\altaffilmark{1}} 
\affil{blink@dante.physics.montana.edu} \affil{Montana State 
University, Department of Physics, Bozeman MT 59717} 
\altaffiltext{1}{Also Los Alamos National Laboratory}

\begin{abstract}

The recent temperature measurements of the two older isolated neutron 
stars PSR 1929+10 and PSR 0950+08 (ages of $3\times 10^6$ and $2\times 
10^7$ yr, respectively) indicate that these objects are heated.  A 
promising candidate heat source is friction between the neutron star 
crust and the superfluid it is thought to contain.  We study the 
effects of superfluid friction on the long-term thermal and rotational 
evolution of a neutron star.  Differential rotation velocities between 
the superfluid and the crust (averaged over the inner crust moment of 
inertia) of $\bar\omega\sim 0.6$ rad s$^{-1}$ for PSR 1929+10 and 
$\sim 0.02$ rad s$^{-1}$ for PSR 0950+08 would account for their 
observed temperatures.  These differential velocities could be 
sustained by pinning of superfluid vortices to the inner crust lattice 
with strengths of $\sim$ 1 MeV per nucleus.  Pinned vortices can creep 
outward through thermal fluctuations or quantum tunneling.  For 
thermally-activated creep, the coupling between the superfluid and 
crust is highly sensitive to temperature.  If pinning maintains large 
differential rotation ($\sim 10$ rad s$^{-1}$), a feedback instability 
could occur in stars younger than $\sim 10^5$ yr causing oscillations 
of the temperature and spin-down rate over a period of $\sim 0.3 
t_{\rm age}$.  For stars older than $\sim 10^6$ yr, however, vortex 
creep occurs through quantum tunneling, and the creep velocity is too 
insensitive to temperature for a thermal-rotational instability to 
occur.  These older stars could be heated through a steady process of 
superfluid friction.

\end{abstract}


\section{Introduction}

A cooling neutron star cools initially through neutrino emission 
before making a transition to photon cooling at an age of $\sim 10^5$ 
yr (see Tsuruta 1998 for a comprehensive review; see Fig.  1).  
Internal heating processes, if they occur, could affect when and how 
abruptly the star makes the transition to photon cooling.  Later, when 
the heating power begins to exceed the luminosity from residual heat, 
the heat source would control the star's thermal evolution.  Internal 
heating processes that could occur include superfluid frictional 
heating (see, \eg, \cite{greenstein75}; \cite{HGG}; Alpar \etal\ 1987; 
Shibazaki \& Lamb 1989; \cite{VEM}; Umeda \etal\ 1993; Van Riper, Link 
\& Epstein 1995), structural readjustment through plastic flow or 
``starquakes'' (\cite{starquakes}; \cite{ruderman76}; Cheng \etal\ 
1992), chemical disequilibrium driven by the star's spin-down 
(Reisenegger 1995), and magnetic field decay (Thompson \& Duncan 1996; 
Heyl \& Kulkarni 1998).  Temperature measurements of neutron stars 
older than $\sim 10^5$ yr provide strong tests of cooling models in 
the photon cooling era, and offer the possibility of constraining the 
heating processes that might occur.  In particular, the recent 
temperature measurements of PSRs 1929+10 and 0950+08 by Pavlov, 
Stringfellow \& C\'ordova (1996) pose a serious challenge to standard 
cooling models, and seem in fact to demand internal heating at rates 
of $\sim 10^{-4} L_\odot$ and $\sim 10^{-5} L_\odot$, respectively 
(see Fig.  1.)

In this paper we explore the possibility that heat generated by 
friction between the neutron star crust and the interior neutron fluid 
is the dominant heating processes taking place in old neutron stars 
with conventional magnetic fields ($\sim 10^{12}$ G).  Most of the 
mass of a neutron star is expected to be in the form of a neutron 
superfluid that condenses shortly after the star's birth (Migdal 
1959).  Large velocity differences between the superfluid and the 
crust could develop in the star's inner crust, where the vortices that 
thread the rotating superfluid pin to nuclei (\cite{AI}; 
\cite{ruderman76}; Alpar 1977; Epstein \&\ Baym 1988; \cite{pizzo}).  
Differential rotation between the stellar crust and the superfluid 
would lead to frictional heat generation, while variations in the 
frictional coupling would affect the star's spin behavior.  Studies of 
superfluid friction in the inner crust indicate that the coupling can 
be highly temperature-dependent, scaling with temperature as ${\rm 
e}^{-E/kT}$, where $E$ is an energy $\gg kT$ (see, \eg, Alpar \etal\ 
1984; Link, Epstein \& Baym 1993; Chau \& Cheng 1993).  The star's 
thermal and rotational evolution are thus coupled and must be 
considered together.  Usually the core superfluid is regarded as 
corotating with the stellar crust, though Sedrakian \& Sedrakian 
(1995; see also Sedrakian \& Cordes 1998) have suggested that 
interactions between superfluid vorticity in the core with the London 
current near the crust-core interface could maintain differential 
rotation between the crust and core.  In this case, friction between 
the core superfluid and the normal matter could also be a heat source.

The goal of this paper is to describe the role played by superfluid 
friction in late-time neutron star thermal evolution.  Though heat 
generation in the core is a possibility, we will focus primarily on 
friction in the crust, as the coupling there has been studied in 
detail.  We consider first (\S 3) the case in which the superfluid is 
in rotational equilibrium, spinning down at the same rate as the 
crust.  This case was originally considered by Alpar \etal\ (1987) and 
applied to the upper limit on the luminosity of PSR 1929+10 available 
at the time to obtain a constraint on the excess angular momentum 
residing in the superfluid.  In later work, Shibazaki \& Lamb (1989), 
Umeda \etal\ (1993), and Van Riper, Link \& Epstein (1995) included 
the effects of superfluid friction in simulations of neutron star 
thermal evolution to obtain further constraints on the heating through 
this process.  This study is largely motivated by the recent 
measurements of the temperatures of the old pulsars PSRs 1929+10 and 
0950+08 (Pavlov, Stringfellow \& C\'ordova 1996), which, as we show, 
provide the most stringent constraints to date on the rotation of the 
superfluid interior.

A crucial issue in neutron star thermal evolution is the possible 
development of thermal instabilities.  Shibazaki and Mochizuki (1994; 
hereafter SM) have shown that under certain circumstances a feedback 
instability between the star's thermal and rotational states can occur 
if the frictional coupling of the superfluid to the crust is 
sufficiently sensitive to temperature; as heat is generated, the 
frictional coupling is increased, creating more heat.  The star 
executes a limit cycle in which its temperature oscillates about the 
temperature at which it is marginally stable, accompanied by 
oscillations in the rotation rate.  This case we study in detail in \S 
4.  We conclude that the coupling of the superfluid to the crust is 
nearly independent of temperature in a star older than $\sim 10^6$ yr, 
effectively decoupling the star's rotational and thermal evolution.  
In \S 5 we use x-ray and optical data from cooling neutron stars to 
obtain constraints on the excess angular momentum residing in the 
superfluid.

\section{Candidate Heating Processes in Old Pulsars}

Fig.  1 shows a comparison of thermal evolution calculations 
(neglecting possible internal heating) with the surface temperature 
measurements of eight neutron stars.  Residual heat is adequate to 
explain the temperatures of the younger objects, but cannot account 
for the temperatures of PSR 1929+10 and PSR 0950+08.  Pavlov, 
Stringfellow \& C\'ordova (1996) have measured temperatures of 
$T_{s}^{\infty} = 1.0 - 3.0 \times 10^{5}$ K for PSR 1929+10 and $6.6 
- 7.4 \times 10^{4}$ K for PSR 0950+08 (see \S 5.1 for further 
discussion of the observations).  The discrepancy between these 
measurements and the predictions of cooling theory is far too large to 
be accounted for by atmospheric uncertainties or modification of the 
energy transport by a magnetic field, and internal heating is 
required.  We now discuss several candidate heating processes: 
superfluid friction, structural relaxation, chemical disequilibrium 
and magnetic field decay.

Differential rotation between the neutron star crust and the neutron 
superfluid would generate heat through friction.  If the superfluid 
and crust are in rotational equilibrium with respect to each other 
(both slowing down at the same rate), the heating power is (Shibazaki 
\& Lamb 1989; Van Riper, Link \& Epstein 1995; Umeda \etal\ 1993; see 
eq.  [\ref{pp}] below)

\begin{equation}
\label{rehe}
	H (t) = {\Delta J_s} \big\vert {\dot\Omega}_0 \big\vert \equiv I_s 
	{\bar\omega} \big\vert {\dot\Omega}_0\big\vert,
\end{equation}
where $\Delta J_s$ is the excess angular momentum residing in the 
superfluid, $I_s$ is the moment of inertia of the portion of the 
superfluid that is differentially rotating, $\bar\omega$ is the 
angular velocity difference between the two components averaged over 
the superfluid moment of inertia and $\vert {\dot\Omega}_0\vert$ is 
the spin-down rate.  In principle, the heat could be generated 
anywhere in the star in which there are superfluid neutrons.  Usually, 
however, the superfluid in the core is regarded as tightly coupled to 
the rotation of the solid through Fermi liquid effects (\cite{ALS}).  
Analyses of glitch data (\cite{AEO}) and spin variations in accreting 
neutron stars (\cite{Boyntonetal}) and in isolated pulsars 
(\cite{boynton}; \cite{deeter}) support this picture.  In the inner 
crust, however, interaction between crustal nuclei and superfluid 
vorticity could lead to substantial differential rotation (see, e.g., 
\cite{AI}) and heating.  The velocity difference that can develop is 
determined by uncertain microphysics, however, $\bar\omega$ could 
exceed $\sim 10$ rad s$^{-1}$ (\cite{EB}; \cite{pizzo}).  In a star 
with a stiff equation of state and a thick crust, $\Delta J_s$ could 
then be as large as $\sim 6\times 10^{45}$ ergs s, in which case the 
heating power is approximately

\begin{equation}
\label{heatmag}
   H (t) = 0.8 \,\left ({\Delta J_s}\over {6\times 10^{45} \ {\rm ergs 
   \ s}} \right ) \left ( {P\over {0.2\ {\rm s}}}\right )^{-1} \left ( 
   { {t_{\rm age}}\over {10^6\ {\rm yr}} }\right )^{-1}\ L_\odot,
\end{equation}

  \noindent where $P$ is the star's spin period, $t_{\rm age}$ is the 
  spindown age $\equiv P/2\dot P$, and a rapid initial spin rate was 
  assumed.  A heating rate this large would begin to play an important 
  role in thermal evolution after $\sim 10^5$ yr, when $H(t)$ becomes 
  comparable to the luminosity from residual heat.

Another process that heats the star is structural relaxation occurring 
as the star spins down and becomes less oblate.  The neutron star 
crust probably becomes brittle when its temperature drops below $\sim 
10^8$ K (Ruderman 1976) at an age of $\sim 10^4$ yr, and subsequently 
suffers structural relaxation through violent starquakes.  The rate of 
heat generation is then of order (Cheng \etal\ 1992),
\begin{equation}
\dot E_{\rm quake} \sim {B \theta_c^2\over t_{\rm age}} = 10^{-5} 
\left ({B\over 10^{48}\ {\rm erg}}\right ) \left ({\theta_c\over 
10^{-3}}\right )^2 \left ({t_{\rm age}\over 10^6\ {\rm yr}}\right 
)^{-1} L_\odot,
\end{equation}
where $B$ is the shear modulus of the crust and $\theta_c$ is the 
critical strain angle at which the crust breaks.  The critical strain 
angle is quite uncertain.  Cheng \etal \ (1992) have shown that 
structural relaxation could constitute an important heat source in 
older stars if $\theta_c$ is $\gap 10^{-2}$.  However, a value of 
$\theta_c$ this large is close to that for a perfect Coulomb lattice, 
while lattice imperfections most likely make $\theta_c$ considerably 
smaller (Smolukowski 1970).  We assume that superfluid friction 
dominates starquake heating.

Reisenegger (1995) has studied heating arising as a neutron star spins 
down and compression of the matter drives it from chemical 
equilibrium.  For the $\sim 10^{12}$ G fields expected for most 
isolated neutron stars this process appears to be relatively 
unimportant, but could be relevant in stars with small magnetic fields 
($\lap 10^{10}$ G).  Reisenegger cautions, however, that chemical 
disequilibrium in superfluid neutron matter could give larger heating 
rates than those he estimates for normal matter.

Thompson \& Duncan (1996) have studied heat generation by the decay of 
a strong magnetic field, and argue that the dominant decay process of 
a $\gap 10^{14}$ G field is the irrotational mode of ambipolar 
diffusion (Goldreich \& Reisenegger 1992).  Equating the rate of 
energy loss by field decay to the photon luminosity gives a surface 
temperature of
\begin{equation}
		4\pi\sigma R^{2}T_{s}^{4} \simeq \left(\frac{4\pi 
		R^{3}}{3}\right)\left(\frac{B_{p}^{2}}{8\pi 
		t_{ambip}^{irr}}\right)
\end{equation}
with
\begin{equation}
		t_{ambip}^{irr} \simeq \frac {5 \times 
		10^{15}}{T_{8}^{6}B_{12}^{2}}{\rm yr}.
\end{equation}
Relating the surface temperature to the internal temperature using the 
results of Gudmundsson, Pethick \& Epstein (1982; see eq.  [\ref{gpe}] 
below) we obtain
\begin{equation}
		T_{s} = 5.8 \times 10^{5}\left(\frac{B} {10^{16}{\rm 
		G}}\right)^{-0.57}.
\end{equation}
The fields required to account for the observed temperatures of PSRs 
1929+10 and 0950+08 are $\sim 10^{16}$ G and $\sim 10^{17}$ G, 
respectively.  By contrast, the {\em dipole} fields of these objects 
inferred from the vacuum dipole model are only $\sim 10^{11}$ G. Thus, 
heating by magnetic field decay appears an unlikely heat source to 
power the emission of PSRs 1929+10 and 0950+08.

\section{Steady Heating From Superfluid Friction}

The first situation we consider is internal heating arising from 
steady slow-down of the neutron superfluid.  In the next section, we 
study perturbations to this equilibrium state.

We describe a neutron star as consisting of two components - a solid 
crust and an interior liquid (the superfluid).  The crust is acted 
upon by an external torque, and the crust and superfluid are coupled 
through friction.  Regardless of the system's initial spin state, 
eventually an equilibrium will be reached in which the two components 
are both spinning down at the same rate, with the liquid spinning more 
rapidly than the solid to a degree determined by the external torque 
and the strength of the frictional coupling.  In this state of 
rotational equilibrium, friction between the two components generates 
heat at a rate that is nearly constant (for an external torque that is 
changing slowly).

The rate of heat production is given by the difference between the 
rate of change of the total rotational energy of the star and the rate 
at which work is done by the external torque (\cite{SL}; Van Riper, 
Link \& Epstein 1995):
\begin{eqnarray}
H (t) &=& N_{ext}\Omega_c(t) 
-\frac{d}{dt}\left[\frac{1}{2}I_c\Omega_c^2(t) + \frac{1}{2} \int 
dI_{s}\,\Omega_s^2({\hbox{\bf r}},t)\right] \nonumber \\
&=& \int dI_{s}\vert\dot\Omega_{s}({\mathbf r},t)\vert\omega({\mathbf 
r},t),
\label{ee}
\end{eqnarray}
where $N_{ext}$ is the external braking torque, $\Omega_c (t)$ is the 
angular velocity of the crust and any components of the star tightly 
coupled to it, $\Omega_s ({\mathbf r},t)$ is the superfluid angular 
velocity at position ${\mathbf r}$ in the star, $\omega({\mathbf r},t) 
\equiv \Omega_s({\mathbf r},t) - \Omega_c(t)$ is the angular velocity 
{\it lag} between the superfluid and the crust, and $I_s$ is the 
superfluid moment of inertia.  Eq.  [\ref{ee}] gives the heating rate 
whether the two components are in rotational equilibrium or not.  In 
rotational equilibrium, the crust superfluid is everywhere spinning 
down at the rate of the crust $\vert\dot\Omega_0 (t)\vert$, and the 
heating rate is
\begin{equation}
H = \Delta J_s |\dot\Omega_0 (t)|,
\label{pp}
\end{equation}
where $\Delta J_s\equiv \int dI_s \omega_0 ({\mathbf r},t)$ is the 
excess angular momentum in the superfluid and $\omega_0 ({\mathbf 
r},t)$ is the lag in equilibrium.  After $\sim 10^6$ yr, when the star 
has lost most of its residual heat, the heating rate is approximately 
balanced by cooling through the emission of surface photons, \ie,

\begin{equation}
4\pi\sigma R_{\infty}^{2}T_{s,\infty}^{4} = \Delta J_s 
|\dot\Omega_{0,\infty} (t)|,
\label{steadyrate}
\end{equation}
where $R_{\infty}$ is the radius, $T_{s,\infty}$ is the surface 
temperature and $|\dot\Omega_{0,\infty} (t)|$ is the spin-down rate 
(subscript $\infty$ indicates quantities seen by a distant observer; 
$\Delta J_s$ is evaluated at the stellar surface).  Observed stellar 
quantities are related to their values at the surface through the 
redshift $e^{-\Phi} \equiv (1 - 2GM/Rc^2)^{-1/2}$ as
\begin{equation}
	T_{s,\infty} = e^{\Phi}T_{s} \qquad |\dot\Omega_{0,\infty} (t)| = 
	e^{2\Phi}|\dot\Omega_{0} (t)| \qquad R_{\infty} = e^{-\Phi}R.
\end{equation}
In \S 5, we will apply eq.  [\ref{steadyrate}] to PSRs 1929+10 and 
0950+08 to obtain estimates for the values of $\Delta J_s$ required to 
heat these sources to their observed temperatures.

\section{Thermal-rotational Instability}

We now study the stability of the rotational equilibrium described 
above by determining how a neutron star containing a pinned inner 
crust superfluid responds to a perturbation of its thermal and 
rotational state.  This problem was originally considered by SM under 
the simplifying assumption that the superfluid angular velocity has no 
gradients.  We extend their work to account for gradients in the 
superfluid angular velocity lag (as would arise from vortex pinning) 
and the effects of quantum tunneling and vortex self-energy on the 
superfluid dynamics.  We find (as did SM) that under some 
circumstances a feedback instability that couples the star's thermal 
and rotational states can occur.  In contrast to SM however, we find 
that the thermal-rotational instability cannot occur in stars older 
than $\sim 10^{6}$ yr.  While our discussion will be formulated with 
coupling to the inner crust superfluid in mind, many of our results 
are general and can be applied to coupling with a core liquid.

\subsection{Perturbation Analysis}

The star's total angular momentum changes under an external torque as 
(neglecting general relativistic effects)
\begin{equation}
\dot J_{\rm tot}(t) = I_c\dot\Omega_c(t) + \int 
dI_s\,\dot\Omega_s({\mathbf r},t) = N_{\rm ext}(t) \equiv 
-I\vert\dot\Omega_0 (t)\vert,
\label{jdot}
\end{equation}
where $I_c$ is the moment of inertia of the crust plus any other 
component(s) to which it is effectively coupled (\eg, the core), $I_s$ 
is the superfluid component in differential rotation, and $I \equiv 
I_c + I_s$ is the total moment of inertia.  In rotational equilibrium, 
the crust superfluid, the crust and the core are all spinning down at 
a rate $\vert\dot\Omega_0(t)\vert$.

A neutron star becomes isothermal within $\sim 10^4$ yr after its 
birth (Van Riper 1991; Umeda \etal\ 1993).  The star's thermal 
evolution is then governed by
\begin{equation}
C_v\dot T(t) = H + \Lambda_{c},
\label{FF}
\end{equation}
where $T$ is the internal temperature, $C_v$ is the heat capacity, $H$ 
is the internal heating rate and $\Lambda_{c}$ is the cooling rate.  
The star cools through emission processes that depend on age and 
composition.  We parameterize the cooling rate as
\begin{equation}
\Lambda_{c} = -BT^n(t),
\label{lambda}
\end{equation}
where $B$ and $n$ are constants which depend on the cooling mechanism.  
For the first $\sim 10^5$ yr of a neutron star's thermal evolution, 
the dominant mode of energy loss is through modified ($n=8$) or direct 
($n=6$) URCA reactions.  Later, the star cools primary through the 
emission of photons from the surface ($n\simeq 2.2$).  Combining eq.  
[\ref{pp}] for the equilibrium heating rate with eqs.  [\ref{FF}] and 
[\ref{lambda}] gives the temperature evolution equation for the 
equilibrium state:
\begin{equation}
C_{v,0} \dot T_0 = - B T_0^n + \Delta J_s\vert\dot{\Omega}_0(t)\vert,
\label{Tss}
\end{equation}
where here and henceforth the subscript ``0'' denotes equilibrium 
quantities, themselves functions of time.

The heating rate in eq.  [\ref{ee}] is determined by the rotational 
state of the superfluid.  The superfluid obeys the equation of motion 
(\cite{bc}),
\begin{equation}
{\partial {\mathbf w}({\mathbf r}, t) \over \partial t} + 
{\mathbf\nabla}\times\left [{\mathbf w}({\mathbf r}, t) \times 
{\mathbf v_v}({\mathbf r}, t)\right ] = 0,
\label{vorticity}
\end{equation}
where ${\mathbf w}\equiv {\mathbf\nabla} \times {\mathbf v_s}$ is 
the superfluid vorticity, ${\mathbf v_s}$ is the fluid velocity and 
${\mathbf v_v}$ is the vortex velocity.  All quantities are averaged 
over regions containing many vortices.  The circulation around any 
contour is given by
\begin{equation}
\oint d{\mathbf l}\cdot {\mathbf v_s} = \int_A d{\mathbf A}\cdot 
{\mathbf  w} = \kappa N,
\label{circ}
\end{equation}
where $N$ is the number of vortices surrounded by the contour of area 
$A$ and $\kappa$ is the quantum of circulation ($h/2m_n$; $m_n=$ 
neutron mass).  The component of the vorticity along the axis of 
rotation is related to the areal density $n$ of vortices in the 
perpendicular plane.  For rotation along the $z$-axis, eq.  
[\ref{circ}] gives
\begin{equation}
{\rm w}_z ({\mathbf r}, t) = \kappa n ({\mathbf r}, t).
\end{equation}
The $z-$component of eq.  [\ref{vorticity}] gives the conservation law
\begin{equation}
{\partial n ({\mathbf r}, t)\over\partial t} + {\mathbf\nabla}\cdot 
\left [n ({\mathbf r}, t){\mathbf v_v}({\mathbf r}, t) \right ] = 0.
\label{continuity}
\end{equation}
We shall focus on axisymmetric superfluid rotation.  In this case the 
$z$-component $\Omega_s$ of the superfluid angular velocity a distance 
$r_p$ from the rotation axis is given by
\begin{equation}
\oint d{\mathbf l}\cdot {\mathbf v_s} = 2\pi r_p^2 \Omega_s({\mathbf 
r}, t) = \kappa \int_0^{r_p} dr_p^\prime\, 2\pi r_p^\prime n ({\mathbf 
r}, t).
\end{equation}
From eq.  [\ref{continuity}], the equation of motion is
\begin{equation}
\dot\Omega_s({\mathbf r},t) = -{w_z\over r_p} v(\omega,T) = 
-v(\omega,T)\left(\frac{2}{r_p} + \frac{\partial} {\partial 
r_p}\right)\Omega_s({\mathbf r},t),
\label{GG}
\end{equation}
where $v(\omega,T)$, the average radial velocity of vortex lines, is 
determined by the microscopic processes that govern the vortex 
mobility.  In the absence of pinning, the superfluid would approximate 
rigid body rotation so that $\partial\Omega_s/\partial r_p\simeq 0$.  
For simplicity, we assume that pinning introduces gradients in 
$\Omega_s$ that are negligible compared to $2\Omega_s/r_p$.

We assume that thermal conduction maintains isothermality and neglect 
gradients in the perturbed temperature.  We examine the response of 
the system to the following {\em axisymmetric} perturbations of its 
thermal and rotational states:
\begin{equation}
\Omega_s({\mathbf r},t) = \Omega_{s,0}({\mathbf r}) + 
\delta\Omega_s({\mathbf r},t)
\end{equation}
\begin{equation}
\Omega_c(t) = \Omega_{c,0} + \delta\Omega_c(t)
\end{equation}
\begin{equation}
\omega({\mathbf r},t) = \omega_0({\mathbf r}) + \delta\omega({\mathbf 
r},t)
\end{equation}
\begin{equation}
T(t) = T_0 + \delta T(t),
\end{equation}
where unperturbed quantities are evaluated at the time of the initial 
perturbation.  Equation (\ref{jdot}) becomes
\begin{equation}
I_c\delta\dot\Omega_c(t) + \int dI_s\,\delta\dot\Omega_s({\mathbf 
r},t) = 0.
\label{aa}
\end{equation}
To linear order in the perturbations, eq.  [\ref{FF}] becomes the 
following integro-differential equation:
\begin{eqnarray}
C_{v,0}\delta\dot T(t) = - C_{v,0}\frac{\dot T_0}{T_0}\delta T(t) &-& 
Bn T^{n-1}_0 \delta T(t) \nonumber \\ &+& \vert\dot{\Omega}_0\vert 
\int dI_{s}\delta\omega ({\mathbf r},t) - \int dI_{s}\omega_0 
({\mathbf r})\delta\dot\Omega_{s}({\mathbf r},t).
\label{bb}
\end{eqnarray}
Equation (\ref{GG}) becomes
\begin{equation}
\delta\dot\Omega_s({\mathbf r},t) = 
-\vert\dot{\Omega}_0\vert\left(\eta_{\omega}\frac{\delta 
\omega({\mathbf r},t)}{\omega_0({\mathbf r})} + \eta_{_T}\frac{\delta 
T(t)}{T_0} + \frac {\delta\Omega_s({\mathbf 
r},t)}{\Omega_{s,0}({\mathbf r})}\right),
\label{cc}
\end{equation}
where
\begin{equation}
\eta_\omega = \frac{\partial\ln v}{\partial\ln\omega} \qquad\qquad 
\eta_{_T} = \frac{\partial\ln v}{\partial\ln T},
\end{equation}
are evaluated in equilibrium and measure the sensitivity of the vortex 
radial velocity to changes in the lag and temperature.  The quantities 
$\eta_\omega$ and $\eta_{_T}$ could have spatial dependence, but for 
the vortex velocity we will adopt below they are nearly constant.

We seek separable solutions for the perturbed quantities of the form 
$\delta A({\mathbf r}, t) = \delta A({\mathbf r})e^{-i\omega t}$.  
Combining eqs.  [\ref{aa}] - [\ref{cc}], we obtain
\begin{eqnarray}
	i\omega \left[\frac{\eta_{\omega}\eta_{_T}}{T_{0}}I_0^{2}\right.  
	&+& \left.  \frac {\eta_{_T}I_{c}}{T_{0}}I_1 - 
	C_{v,0}\eta_{\omega}I_{-1} - 
	\frac{\eta_{\omega}\eta_{_T}}{T_{0}}I_1I_{-1} + 
	I_{c}C_{v,0}\right] - C_{v,0}I_{c} {\dot T_{0}\over T_0} \nonumber 
	\\
	&-& BnT_{0}^{n-1}I_{c} + \frac {\vert\dot\Omega_{0}\vert I 
	\eta_{_T}}{T_{0}}I_0 + \frac{C_{v,0}\dot 
	T_{0}\eta_{\omega}}{T_{0}}I_{-1} + 
	BnT_{0}^{n-1}\eta_{\omega}I_{-1} = 0,
	\label{intdisp}
\end{eqnarray}
where
\begin{equation}
	I_n \equiv \int dI_s \omega_0^n 
	\left(\frac{i\omega}{\vert\dot\Omega_{0}\vert} - 
	\frac{\eta_{\omega}}{\omega_{0}({\mathbf r})} - 
	\frac{1}{\Omega_{s,0}({\mathbf r})}\right)^{-1}
\end{equation}
Evaluation of eq.  [\ref{intdisp}] requires a form for 
$\omega_{0}({\mathbf r})$.  Pinning calculations indicate that the 
pinning energy per nucleus peaks near a density of $\sim 10^{14}$ g 
cm$^{-3}$ (\cite{EB}; \cite{pizzo}); in these regions, $\omega_0$ will 
be the largest.  To model this behavior, we treat $\omega_0$ as taking 
a value $\omega_0=\omega_{0,{\rm max}}$ through a region of total 
moment of inertia $\Delta I_s$, and zero otherwise.  Since the pinning 
energy is largest in the densest regions of the crust, $\Delta I_{s}$ 
probably accounts for most of the crust's moment of inertia.  
Evaluating the integrals in eq.  [\ref{intdisp}], we find,
\begin{equation}
	\omega^{2} + i\omega Y - Z = 0,
\label{mm}
\end{equation}
with
\begin{equation}
	Y = \frac{\dot T_0}{T_{0}} + \frac{B n T_0^{n-1}}{C_{v,0}} + 
	\tilde \Omega - \frac{\vert\dot\Omega_{s,0}\vert \eta_{_T} 
	\omega_{0,{\rm max}}\Delta I_{s}}{C_{v,0} T_{0}},
\label{OO}
\end{equation}
\begin{equation}
Z = \left ({\dot{T_{0}}\over T_0} + {B n T_0^{n-1}\over C_{v,0}} 
\right ) \tilde{\Omega} + {\eta_{_T}\vert\dot\Omega_{0}^{2}\vert I 
\Delta I_{s}\over C_{v,0}T_{0}I_{c}},
\end{equation}
where
\begin{equation}
\tilde \Omega \equiv \vert\dot\Omega_{s,0}\vert 
\left(\frac{1}{\Omega_{s,0}}+\frac{(I_{c}+\Delta 
I_{s})\eta_\omega}{I_c\omega_{0,{\rm max}}}\right).
\end{equation}
Perturbations are damped when $Y>0$, unstable when $Y<0$, and 
marginally stable (or oscillatory) when $Y = 0$.  The condition for a 
star to be unstable is,
\begin{equation}
\frac{\dot T_0}{T_{0}} + \frac{B n T_0^{n-1}}{C_{v,0}} + \tilde \Omega 
- \frac{\vert\dot\Omega_{s,0}\vert \eta_{_T} \omega_{0,{\rm 
max}}\Delta I_{s}}{C_{v,0} T_{0}} \leq 0.
\end{equation}
Using the equilibrium equation, eq.  [\ref{Tss}], we make the 
replacement
\begin{equation}
BT_0^{n} = \vert\dot\Omega_{0}\vert\Delta I_s \omega_{0,{\rm max}} - 
C_{v,0} \dot T_0,
\end{equation}
to obtain a quadratic equation for the critical temperature $T_c$,
\begin{equation}
		T_{c}^{2} + \frac {(n-1)I_{c}\omega_{0,{\rm max}}\vert\dot 
		T\vert}{ \eta_\omega I \vert\dot\Omega_{0}\vert}T_{c} + \frac 
		{(n-\eta_{_T})I_{c}\omega_{0,{\rm max}}^{2} \Delta I_{s}}{a I 
		\eta_\omega} = 0,
\label{quad}
\end{equation}
where $a\equiv C_{v,0}/T_0$ is independent of temperature for 
degenerate matter.  We have assumed $\tilde\Omega \sim 
\vert\dot\Omega_{0}\vert\eta_{\omega}I/\omega_{0,{\rm max}}I_{c}$, 
which is valid as long as $\omega_{0,{\rm max}} \ll 
\eta_\omega\Omega_{s}$.  Eq.  [\ref{quad}] has a positive solution:
\begin{equation}
		T_{c} = - \frac {(n-1)\omega_{0,{\rm max}}I_{c}\vert\dot 
		T\vert}{ 2\eta_\omega I \vert\dot\Omega_{0}\vert} + 
		\frac{1}{2}\left(\frac {(n-1)^{2}\omega_{0,{\rm 
		max}}^{2}I_{c}^{2}\vert\dot T\vert^{2}}{\eta_\omega^{2}I^{2} 
		\vert\dot\Omega_{0}\vert^{2}} + 
		\frac{4(\eta_{_T}-n)\omega_{0,{\rm max}}^{2}I_{c} \Delta 
		I_{s}}{a \eta_\omega I}\right)^{1/2}.
\end{equation}
Below we estimate $\eta_{_T} = \eta_\omega \simeq 30$.  For a star of 
age $\sim 10^4$ yr, $\vert\dot T\vert$ is $\sim 10^{-7}{\rm K \ 
s^{-1}}$ and $\vert\dot\Omega_{0}\vert$ is $\sim 10^{-10}{\rm rad \
s^{-2}}$.  Taking $\omega_{0,{\rm max}}\sim 10$ rad s$^{-1}$ and $n = 
8$ (modified URCA cooling), we see that the first term under the 
square root is negligible compared to the second.  The critical 
temperature is thus approximately
\begin{equation}
		T_c^2 \simeq \frac{(\eta_{_T} - n)\omega_{0,{\rm 
		max}}^{2}I_{c}\Delta I_{s}}{a\eta_\omega I}.
\label{JJ}
\end{equation}
From eq.  [\ref{JJ}], we see that a minimum sensitivity of the vortex 
velocity to temperature is required for a thermal-rotational 
instability to occur; for $\eta_{_T}<n$, the star is stable at any 
temperature.  Eq.  [\ref{JJ}] agrees with eq.  [24] of SM in the limit 
$\Delta I_s=I_s$, $\eta_{_T} \gg n$ and $I_{c} \simeq I$.

Eq.  [\ref{JJ}] can be applied for any coupling of the superfluid and 
crust that depends on $T$ and $\omega$.  In principle, the superfluid 
component with excess angular momentum $\omega_{0,{\rm max}}\Delta 
I_s$ could be anywhere in the star.  In the inner crust, however, 
pinning of vortices to the nuclear lattice (\cite{AI}; 
\cite{ruderman76}; Alpar 1977; Epstein \& Baym 1988; \cite{pizzo}) 
could sustain significant differential rotation between the superfluid 
and the crust.  The mobility of vortices in the presence of pinning 
determines the superfluid's ability to respond to changes in 
temperature and rotation rate.  Vortex dynamics in the presence of 
pinning has been studied in detail by Link \& Epstein (1991; hereafter 
LE) and LEB under the assumption that vortex stresses do not break the 
nuclear lattice.  We now review the key results of this work, and 
study the implications of vortex pinning for the thermal-rotational 
instability.

\subsection{Vortex Dynamics in the Presence of Pinning}

Were superfluid vortices perfectly pinned to the inner crust lattice, 
the superfluid velocity would be fixed.  As the solid crust slows 
under the external torque, a velocity difference between the crust and 
superfluid would develop exerting a Magnus force on the pinned 
vortices directed radially outward.  If the lag $\omega$ locally 
exceeds a critical value $\omega_c$, vortices cannot remain pinned in 
the presence of the Magnus force; the vortices unpin and flow outward, 
spinning down the superfluid.  The critical lag is determined by the 
condition that the Magnus force per unit length of vortex equal the 
pinning force per unit length (see, \eg, LE)
\begin{equation}
{F_p\over l} = \rho_{s}\kappa r_p \omega_c,
\label{pinforce}
\end{equation}
where $F_p$ is the pinning force per nucleus, $l$ is the lattice 
spacing and $\rho_{s}$ is the superfluid mass density.

For $\omega<\omega_c$, pinned vortices can still move outward as 
thermal or quantum excitations allow them to overcome their pinning 
barriers.  The resulting average velocity of {\em vortex creep} is 
determined by the pinning strength, the properties of vortices, the 
characteristic energy of excitations on a pinned vortex, and the 
velocity difference between a pinned vortex and the superfluid flowing 
past it.  Accounting for quantum effects and the vortex self-energy, 
LEB and LE obtain a creep velocity of the form (see eq.  6.9 of LEB)
\begin{equation}
	v_{cr} = v_0 \exp (- A (\omega)/T_{\rm eff}).
\label{ddd}
\end{equation}
Here $A$ is the {\em activation energy} for a segment of vortex line 
to overcome its pinning barrier; it decreases with $\omega$, becoming 
zero for $\omega=\omega_c$.  The prefactor $v_{0}$ is a microscopic 
velocity comparable to the radial component of the velocity of an 
unpinned vortex segment.  The radial velocity is determined by the 
dissipative processes associated with vortex motion.  Epstein \& Baym 
(1992) have shown that drag arising from the excitation of {\em Kelvin 
modes} causes free vortices to move radially outward at velocities 
comparable to the velocity difference between the superfluid and 
normal matter.  For the pinning energies estimated by Epstein \& Baym 
(1988), this velocity difference could be $\sim 10^6$ cm s$^{-1}$; we 
estimate $v_0\simeq 10^6$ cm s$^{-1}$.  The effective temperature, 
$T_{\rm{eff}}$, is (eq.  4.10, LEB)
\begin{equation}
T_{\rm eff} = T_q {\rm coth} {T_q\over T},
\end{equation}
where $T_q$ is the {\em cross-over temperature} that determines the 
transition from vortex motion through thermal activation to quantum 
tunneling.  For $T\gg T_q$, vortices move primarily through classical 
thermal activation.  For $T\ll T_q$, the dominant process is quantum 
tunneling.  In these two limits
\begin{equation}
T_{\rm eff} \rightarrow \left\{ \begin{array}{ll} T & \mbox{for $T\gg 
T_q$ (classical)} \\
						T_q & \mbox{for $T\ll T_q$ (quantum)}
						\end{array}
						\right.
\end{equation}

In equilibrium, the superfluid and the crust are both spinning down at 
a rate $\vert\dot\Omega_0(t)\vert$.  From eqs.  [\ref{GG}] and 
[\ref{ddd}], the equilibrium state satisfies
\begin{equation}
{A(\omega_0)\over T_{\rm eff}} = \ln {4 v_0 t_{\rm age}\over r_p},
\label{steadystate}
\end{equation}
where we took $\Omega_s/\vert\dot\Omega_0\vert\simeq 2 t_{\rm age}$.  
The pinning force is a function of density alone, and so is constant 
on spherical shells.  From eq.  [\ref{steadystate}], we see that 
$\omega_0$ is nearly constant on such shells except near the 
rotational poles.  For purposes of obtaining estimates, we henceforth 
take $r_p\simeq R$.  For $v_0=10^6$ cm s$^{-1}$ and $t_{\rm age}=10^6$ 
yr,
\begin{equation}
\ln {4 v_0 t_{\rm age}\over R} \simeq 30.
\label{sslag}
\end{equation}
In equilibrium, $A(\omega_0)$ must take a particular local value for 
given $T_{\rm eff}$ and $t_{\rm age}$.  If the pinning force per 
nucleus is relatively small, $\omega_0$ must be $\ll\omega_c$ to 
ensure that $A(\omega)$ is sufficiently large.  On the other hand, if 
the pinning force per nucleus is relatively large, $\omega_0$ must be 
close to $\omega_c$ to allow the vortices to overcome their pinning 
barriers.  We will refer to these two limiting cases as {\em strong} 
and {\em weak pinning}, respectively.  [These cases correspond to the 
limits of {\em flexible} and {\em stiff} vortices discussed by LE].  
In the equilibrium state we have assumed, the local lag has everywhere 
adjusted to the value required to satisfy eq.  [\ref{steadystate}].

Evaluated about equilibrium, the sensitivity of the vortex velocity in 
eq.  [\ref{ddd}] to temperature is
\begin{equation}
\eta_{_T} = \left [\ln {4 v_0 t_{\rm age}\over R}\right ] \left 
({T_q\over T}\right )^2 {\rm csch}^2\,{T_q\over T}.
\end{equation}
In the classical limit ($T\gg T_q$), $\eta_{_T}\simeq 30$.  Below 
$T=T_q$, both $\eta_{_T}$ and $T_c$ quickly drop to zero as $T$ is 
reduced.  Hence, for the limit cycle to be relevant, $T\gap T_q$ is 
required.

The cross-over temperature $T_q$ is equal to half the ground state 
energy of excitations on a pinned vortex line, and depends sensitively 
on the pinning energy and density.  For weakly-pinned vortices, LEB 
estimate (eq.  3.11, LEB)
\begin{equation}
T_q \simeq 0.2 \left ({\Lambda\over 3}\right )\left ({l\over 50\ {\rm 
fm}}\right )^{-2}\ {\rm keV},
\label{Tqweak}
\end{equation}
where $\Lambda$ is a weak function of density and we have chosen 
fiducial values appropriate to the denser regions of the inner crust.  
At lower density, \eg, near the neutron drip density, $\Lambda$ is 
$\simeq 7$.
 
For strongly-pinned vortices, pinning drives the ground state energy 
of vortex excitations up to a considerably higher value (eq.  3.13, 
LEB):
\begin{equation}
T_q \simeq 60 \left ({\Lambda\over 3}\right )\left ({l\over 50\ {\rm 
fm}}\right )^{-2}\ {\rm keV}.
\end{equation}
To compare these temperatures to those of cooling neutron stars, we 
convert from surface temperature to internal temperature using the 
results of Gudmundsson, Pethick \& Epstein (1982):
\begin{equation}
T_8 = 1.288\left(\frac{T^4_{s,6}}{g_{s,14}}\right)^{0.455}
\label{gpe}
\end{equation}
where
\begin{equation}
g_s = \frac{GM}{R^{2}}e^{-\Phi}.
\end{equation}
Here $T_8 \equiv T/10^8$, $T_{s,6} \equiv T_s/10^6$, $g_{s,14} \equiv 
g_s/10^{14}$ is the surface gravity, and $T_{s,\infty} = 
e^{\Phi}T_{s}$.  For typical neutron star parameters, the internal 
temperature is
\begin{equation}
T = 12\ T_{s,6,\infty}^{1.82}\ {\rm keV}.
\end{equation}
For PSR 0950+08, we estimate 0.09 keV $<T<$ 0.11 keV. This object is 
thus well into the quantum creep regime, for which $\eta_{_T}\simeq 
0$.  The limit cycle cannot occur, and steady slowdown of the crust 
and superfluid is a stable state.  Our conclusion regarding the 
stability of old stars differs from that of SM, who assumed that 
thermal creep is always the dominant process.

For PSR 1929+10 the case is less clear; we find 0.18 keV $<T<$ 1.3 
keV, compared to $T_q\simeq 0.2$ keV estimated in eq.  [\ref{Tqweak}].  
However, this estimate applies only to the special case of extremely 
weak pinning; $T_q$ could be substantially higher than $0.2$ keV. It 
thus appears that PSR 1929+10 is also in the quantum creep regime (or 
borderline), undergoing steady slow-down.

Stars younger than PSR 1929+10 ($t_{\rm age}=3\times 10^6$ yr), could 
be in the thermal creep regime, and hence could be subject to the 
thermal-rotational instability.  Such a star becomes unstable if the 
temperature falls below the critical temperature given by eq.  
[\ref{JJ}].  The quantity $\eta_\omega$, which measures the 
sensitivity of the creep rate to changes in $\omega$ is, from eqs.  
[\ref{ddd}] and [\ref{steadystate}],
\begin{equation}
\eta_\omega = - \left [\ln {4 v_0 t_{\rm age}\over R}\right ] {d\ln 
A\over d\ln\omega}.
\end{equation}
The form for $A(\omega)$ depends on the strength of pinning.  In the 
weak and strong pinning limits, the activation energy is (eq.  B.12, 
LE, in the limit $\omega\ll\omega_c$; eq.  3.15, LE)
\begin{equation}
A (\omega) \rightarrow \left\{ \begin{array}{ll} {\beta\over\omega} & 
\mbox{weak pinning} \\
			U_0 (1 - {\omega\over\omega_c})^{3/2} & \mbox{strong 
			pinning},
						\end{array}
						\right.
\label{activation}
\end{equation}
where $U_0$ is the pinning energy per nucleus, and $\beta$ is a 
parameter that measures the strength of the coupling and is related to 
$U_0$.  For strong pinning, $U_0$ is relatively large ($\gap$ 1 MeV), 
and $T_q$ is up to $\sim 60$ keV. A neutron star is expected to cool 
below this temperature within $\sim 100$ years of its birth.  If 
strong pinning occurs, therefore, quantum tunneling is the dominant 
creep process during most of the star's thermal evolution, and the 
thermal-rotational instability cannot occur.  Hence, the weak pinning 
case is the relevant one for the thermal-rotational instability.  In 
this case
\begin{equation}
\eta_\omega = \ln {4 v_0 t_{\rm age}\over R} \simeq 30.
\end{equation}
Taking $\eta_{_T}=\eta_\omega\simeq 30$, we estimate the internal 
temperature at which the star becomes unstable from eq.  [\ref{JJ}].  
In the limit $I_{s}\ll I_{c}$, appropriate if it is the inner crust 
superfluid that drives the instability, we obtain
\begin{equation}
T_c = 11 \left ({a\over 3.3\times 10^{29}\mbox{ erg s$^{-1}$}}\right 
)^{-1/2} \left ({\omega_{0,{\rm max}}\over 10\mbox{ rad s$^{-1}$}} 
\right ) \left ({\Delta I_s\over 7.3\times 10^{43} \mbox{ g 
cm$^2$}}\right )^{1/2} \,{\rm keV},
\label{Tcest}
\end{equation}
where $n=8$ for modified URCA cooling in a young neutron star.  Here 
and in the following, we estimate stellar parameters such as $I_s$, 
$R$ and $a$ using a $1.4 M_\odot$ stellar model based on the Friedman 
\& Pandharipande (1981) equation of state (see Table 1).  A cooling 
neutron star reaches a temperature of $\sim 10$ keV after $\sim 10^4$ 
yr.  While the critical temperature depends sensitively on the 
uncertain pinning parameters $\omega_{0,{\rm max}}$ and $\Delta I_s$, 
this estimate suggests that young, cooling neutron stars {\em could} 
be unstable to perturbations in temperature and rotation rate.  If, on 
the other hand, $\omega_{0,{\rm max}}$ or $\Delta I_s$ are 
significantly smaller than estimated in eq.  [\ref{Tcest}], the star 
will cool into the quantum creep regime before it can become unstable.

As a star cools and reaches its critical temperature, temperature 
perturbations begin to grow.  Stability is restored as the star is 
heated to slightly above its critical temperature.  The star is again 
able to cool, eventually becoming unstable again.  A limit cycle 
ensues, wherein the star oscillates about its $T_c$ as originally 
demonstrated by SM. To evaluate the characteristic period of the 
oscillations about this marginally-stable state, we solve eq.  
[\ref{mm}] with $Y = 0$.  Using eq.  [\ref{Tss}] and assuming $I_{s} 
\ll I_{c}$, we obtain
\begin{equation}
\tau_{osc} \simeq 
2\pi\left[\frac{\eta_{_{T}}(1+\eta_{\omega})(n-1)BT_{0}^{n-1} 
\vert\dot\Omega_{0}\vert}{a T_{0}(\eta_{_T}-1)\omega_{0,{\rm 
max}}}+\frac{\eta_{\omega}(\eta_{_T}+\eta_{\omega}) 
\vert\dot\Omega_{0}\vert^{2}}{(\eta_{_T}-1)\omega_{0,{\rm 
max}}^{2}}\right]^{-1/2}.
\label{tosc}
\end{equation}
For a young star cooling through the modified URCA process, the first 
term is negligible for $\omega_{0,{\rm max}}\simeq 1-10$ rad s$^{-1}$.  
Taking $\eta_{_T}=\eta_\omega=30$, $\Omega_{s,0}/|\dot\Omega_0| \simeq 
2 t_{age}$ and the spin period as $P\simeq 2 \pi/\Omega_{s,0}$ gives
\begin{equation}
\tau_{osc} \simeq 0.26 \left({P\over 0.1\ {\rm s}}\right) \left( 
{\omega_{0,{\rm max}}\over 10 \ {\rm rad \ s^{-1}}}\right) t_{age}.
\end{equation}

\section{Constraints from Surface Temperature Measurements}

\subsection{Old Pulsars}

Pavlov, Stringfellow \& C\'ordova (1996) have recently detected 
thermal emission from PSRs 1929+10 and 0950+08 in the UV-optical band 
using the COSTAR corrected Faint Object Camera on the Hubble Space 
Telescope.  Assuming the observed flux arises from the entire surface 
of a neutron star with radius 10 km, they obtain surface temperatures 
of $T_{s}^{\infty} = 1.0 - 3.0 \times 10^{5}$ K for PSR 1929+10 and 
$6.6 - 7.4 \times 10^{4}$ K for PSR 0950+08.  Previous X-ray 
observations of PSRs 1929+10 (Yancopoulos, Hamilton, \& Helfand, 1994) 
and 0950+08 (Manning \& Willmore, 1994) produced blackbody fits for 
emitting regions of only $\sim$ 20-30 meters in diameter, suggesting 
that the observed X-ray emission originates from a hot polar cap.  
However, for both of these objects, extension of the blackbody spectra 
into the UV-optical range predicts a flux which is several orders of 
magnitude smaller than that observed by Pavlov, Stringfellow \& 
C\'ordova (1996), consistent with the interpretation that the 
UV-optical emission originates from the entire neutron star surface.

The analysis in \S 4 indicates that these two sources are too cold for 
the thermal-rotational instability to occur.  Assuming steady heating 
by superfluid friction, we determine from eq.  [\ref{steadyrate}] the 
values of the excess angular momentum $\Delta J_s$ and average lag 
$\overline\omega\equiv\Delta J_s/I_s$ required to heat these objects 
to their observed temperatures.  We find $\Delta J_{s} \sim 4 \times 
10^{43}$ ergs s and $\overline\omega \sim 0.6$ rad s$^{-1}$ for PSR 
1929+10, and $\Delta J_{s} \sim 1 \times 10^{42}$ ergs s and 
$\overline\omega \sim 0.02$ rad s$^{-1}$ for PSR 0950+08 (see Table 
2).  Our estimates were obtained for an FP equation of state, which 
gives a radius $R\simeq 11$ km, close to that assumed for the surface 
temperature determinations.  Our results are consistent with earlier 
upper limits on $\bar\omega$ obtained for PSR 1929+10.  Shibazaki \& 
Lamb (1989) obtained $\bar\omega < 0.02$ rad s$^{-1}$ and $\bar\omega 
< 4.0$ rad s$^{-1}$ for stiff and soft equations of state.  Alpar 
\etal\ (1987) found $\bar\omega < 0.7$ rad s$^{-1}$ for a moderate 
equation of state.

Some amount of internal heating also appears to be required for PSR 
1055-52 (see Fig.  1).  \"Ogelman \& Finley (1993) obtained a 
temperature of $T_{s}^{\infty} = 6.9 - 8.1 \times 10^{5}$ K for this 
pulsar using ROSAT PSPC data.  This temperature was obtained by 
interpreting the soft blackbody component of the spectrum as 
originating from a cooling neutron star of radius $\simeq$ 10 km.  If 
we assume that steady heating by the internal superfluid provides the 
heat for this pulsar, we obtain $\Delta J_{s} \sim 5 \times 10^{44}$ 
ergs s and $\overline\omega \sim 7$ rad s$^{-1}$.

\subsection{Young Pulsars}

Surface temperature measurements and upper limits for young pulsars 
are given in Table 3, all fits are blackbody fits with a stellar 
radius of 10 km.  These pulsars could be subject to the 
thermal-rotational instability discussed in \S 4.  We consider PSR 
1055-52 along with the younger pulsars because its internal 
temperature is high enough that the limit cycle cannot be ruled out.  
For most neutron stars younger than $\sim 10^{6}$ yr, the observed 
temperatures can be accounted for by their residual heat content.  As 
discussed in \S 4, if the thermal-rotational instability occurs, a 
cooling neutron star cannot cool below its critical temperature.  By 
requiring the internal temperature deduced from the observed surface 
temperature to be greater than or equal to the critical temperature, 
we obtain from eq.  [\ref{JJ}] the constraint,
\begin{equation}
4.1 \times 10^{4}\left(\frac{(\eta_{_T} - n) \omega_{0,{\rm 
max}}^{2}\Delta I_{s}}{a \eta_\omega R^{1.82} e^{-2.73\Phi}} 
\right)^{0.275} \leq T_{s,\infty},
\label{fff}
\end{equation}
giving an upper limit on $\omega_{0,{\rm max}}^{2}\Delta I_{s}$ 
inasmuch as $R$, $\Phi$, $\eta_{_T}$, $a$ and $n$ are known.

Superfluid pinning is expected to be strongest in the densest regions 
of the inner crust; the characteristic lag $\omega_{0}({\mathbf r})$ 
will be largest in these regions.  It is reasonable to expect then 
that $\omega_{0,{\rm max}}>\bar\omega\equiv I_s^{-1}\Delta J_s$.  For 
this situation, we obtain the constraint,
\begin{equation}
\overline\omega \leq \left(\frac {\omega_{0,{\rm max}}^{2}\Delta 
I_{s}}{I_{s}}\right)^{1/2}.
\label{ooo}
\end{equation}
Where $\omega_{0,{\rm max}}^{2}\Delta I_{s}$ is obtained with eq.  
[\ref {fff}].  For stars younger than $\sim 10^5$ yr, the dominant 
cooling process is neutrino emission; we assume $n=8$ (modified URCA 
process).  We take $\eta_{_T} = \eta_{\omega}=30$, as estimated in \S 
4, and the stellar parameters of Table 1 for $R$, $\Phi$ and $a$.  In 
Table 3 we list constraints from eqs.  [\ref{fff}] and [\ref{ooo}] for 
young neutron stars.  We obtain upper limits on $\bar\omega$ of $\sim 
30$ rad s$^{-1}$, typically.  Note that these constraints apply {\em 
only} if the dominant creep process is classical thermal activation; 
if quantum tunneling is the dominant process, the star is stable at 
any temperature.

\section{Discussion and Conclusions}

PSRs 1929+10 and 0950+08 require significant internal heating to 
account for their observed temperatures.  A promising candidate heat 
source is friction between the neutron star crust and the superfluid 
it contains.  In this paper we have studied the effects of superfluid 
friction on the long-term thermal and rotational evolution of a 
neutron star.  We conclude that average differential rotation between 
the superfluid and the crust of $\bar\omega\sim 0.6$ rad s$^{-1}$ and 
$\sim 0.02$ rad s$^{-1}$ would account for the temperatures of PSRs 
1929+10 and 0950+08 respectively.  A larger lag, $\bar\omega\sim 7$ 
rad s$^{-1}$, is compatible with the temperature of PSR 1055-52.  
These differential velocities could be sustained by the pinning of 
superfluid vortices to the inner crust lattice.

Pinned vortices can creep outward through thermal fluctuations or 
quantum tunneling, depending on the pinning strength and stellar 
temperature.  For thermally-activated creep, the coupling between the 
superfluid and crust is highly sensitive to temperature.  Under some 
circumstances, a feedback instability can occur that brings the 
superfluid and crust closer to corotation and heats the star until 
stability is restored.  A hysteresis develops in which the star 
oscillates about its critical temperature.  For stars older than $\sim 
10^6$ yr, however, vortex creep occurs through quantum tunneling, and 
the creep velocity is too insensitive to temperature for a 
thermal-rotational instability to occur; these stars are stable.  Our 
conclusion regarding the stability of old stars differs from that of 
Shibazaki \& Mochizuki (1994), who assumed that thermal creep is 
always the dominant process.  The thermal-rotational instability 
could, however, occur in younger stars.  Assuming that young stars are 
stable or marginally stable leads to upper limits on the superfluid 
differential velocity of $\sim 10$ rad s$^{-1}$.  These upper limits 
are consistent with the estimates for $\bar\omega$ obtained for the 
older PSRs 1929+10 and 0950+08.

The estimates we obtain for $\bar\omega$ are consistent with 
first-principles calculations of the maximum lag sustainable by 
vortices before unpinning.  Based on the pinning calculations of 
Alpar, Cheng \& Pines (1989) and Ainsworth, Pines \& Wambach (1989), 
Van Riper, Link \& Epstein (1995) obtain an upper limit to the average 
lag velocity of $\bar\omega\sim 10$ rad s$^{-1}$.  From the pinning 
calculations of Epstein \& Baym (1988), Van Riper, Link \& Epstein 
(1995) obtain an upper limit of $\bar\omega\sim 10^2$ rad s$^{-1}$.  
The recent calculations of Pizzochero, Viverit \& Broglia (1997) give 
a pinning force of $F_p=0.63$ MeV fm$^{-1}$ at a density 
$\rho_s=8\times 10^{13}$ g cm$^{-3}$, and a pinning energy of 7.5 MeV; 
the corresponding critical lag is $\omega_c=16$ rad s$^{-1}$.

To estimate the pinning strength required to sustain differential 
rotation of the magnitudes estimated above, we take $F_p$ appearing in 
eq.  [\ref{pinforce}] to be $U_0/r_0$, where $r_0$ is the effective 
range of the pinning potential.  With the lag from PSR 1929+10 we 
obtain,
\begin{equation}
	U_{0} \gap 0.5\, {\rm MeV} \left(\frac{\bar\omega}{0.6\ {\rm rad \ 
	s^{-1}}}\right)\left(\frac{\rho_{s}}{10^{14} \ {\rm g \ 
	cm^{-3}}}\right)\left(\frac{R}{10\ {\rm 
	km}}\right)\left(\frac{r_{0}}{10\ {\rm 
	fm}}\right)\left(\frac{l}{50\ {\rm fm}}\right).
\label{pinenergy}
\end{equation}
Eq.  [\ref{pinenergy}] represents a lower limit since $\bar\omega < 
\bar{\omega}_{c}$, however, $\bar\omega$ could be close to 
$\bar{\omega}_c$ for pinning strengths this large (LEB).  Our 
estimates of $\bar\omega$ for PSR 0950+08 imply a pinning energy 
$\gap$ 0.02 MeV.

The thermal-rotational instability described in this paper might 
produce oscillations in the temperature and spin-down rate of younger 
pulsars with a characteristic period of $\sim 0.3\ t_{\rm age}$.  
Oscillations in the Crab pulsar, for example, could occur over a 
timescale of $\tau_{osc}\sim 10^2$ yr.  Detection of such long-period 
oscillations, especially in the presence of timing irregularities 
(\eg, glitches), would be problematic.  Observational evidence for the 
thermal-rotational instability in any neutron star would offer 
valuable insight into the manner in which the neutron star crust is 
coupled to its superfluid interior.

Our estimates for the excess angular momentum $\Delta J_s$ required to 
heat PSRs 1929+10 and 0950+08 to their observed temperatures hold 
whether the frictional heat is generated in the crust or in the core.  
In obtaining constraints on the average lag $\bar\omega$, we assumed 
coupling to the inner crust superfluid.  If the coupling is elsewhere, 
these estimates scale as $\Delta J_s/I_s$, where $I_s$ is the moment 
of inertia of the component that possesses differential rotation.  The 
upper limits on $\Delta I_s\omega^2_{0,{\rm max}}$ obtained for young 
pulsars apply for the crust since the coupling parameters $\eta_{_T}$ 
and $\eta_\omega$ were determined for vortex creep.

An issue that complicates all interpretations of surface temperatures 
from cooling neutron stars is the uncertainty in atmospheric 
composition.  We used the results of blackbody fits in our analysis.  
Temperature measurements are also available for several pulsars using 
model atmospheres.  Heavy-element atmospheric models produce 
temperatures similar to the blackbody results.  However, non-magnetic, 
light-element atmospheres can give temperatures up to three times 
lower than those obtained with blackbody fits (Romani 1987).  If the 
temperature were in fact three times lower than the blackbody value, 
$\overline\omega$ estimated for PSRs 1929+10, 0950+08 and 1055-52 
would decrease by almost two orders of magnitude.  The upper limits on 
$\overline\omega$ obtained for younger stars would decrease by a 
factor of three.  Hence, the results presented in Tables 2 and 3 are 
conservative and could decrease with improved atmospheric 
considerations and temperature measurements.

Sudden increases in pulsar rotation rates ({\em glitches}) have been 
observed in many younger pulsars and are thought to represent angular 
momentum transfer from the superfluid to the crust.  In the Vela 
pulsar, for example, fractional changes in the rotation rate of the 
crust of $\sim 10^{-6}$ are observed every few years (\cite{CDKP}).  
The maximum angular momentum available for a glitch is $\Delta J_s$, 
giving a maximum glitch magnitude of
\begin{equation}
\frac{\Delta \Omega_{c}}{\Omega_{c}} \simeq \frac{\Delta 
J_s}{I_{c}\Omega_{c}} \simeq 2.6 \times 10^{-2}\left(\frac{\Delta 
J_s}{2 \times 10^{45}\ {\rm ergs\ s}}\right) \left(\frac{I_{c}}{1.1 
\times 10^{45}\ {\rm g \ cm^{2}}}\right)^{-1} 
\left(\frac{\Omega_{c}}{70 \ {\rm s^{-1}}}\right)^{-1}.
\end{equation}
The upper limit of $\bar\omega\simeq 30$ rad s$^{-1}$ obtained for 
Vela gives $\Delta J_s\simeq 2\times 10^{45}$ ergs s for the inner 
crust angular momentum excess, easily compatible with the angular 
momentum requirements of glitches.  The smaller values obtained for 
PSRs 1929+10 and 0950+08 are also adequate to produce Vela-sized 
glitches, though none has been observed.

\acknowledgements

We thank K. Van Riper for providing us with the results of cooling 
simulations and G. Pavlov and D. Page for helpful discussions.  MBL 
would like to thank the Patricia Roberts Harris Graduate Fellowship 
for support.  This work was supported by NASA EPSCoR grant \#291748.

\begin{figure}
\plotone{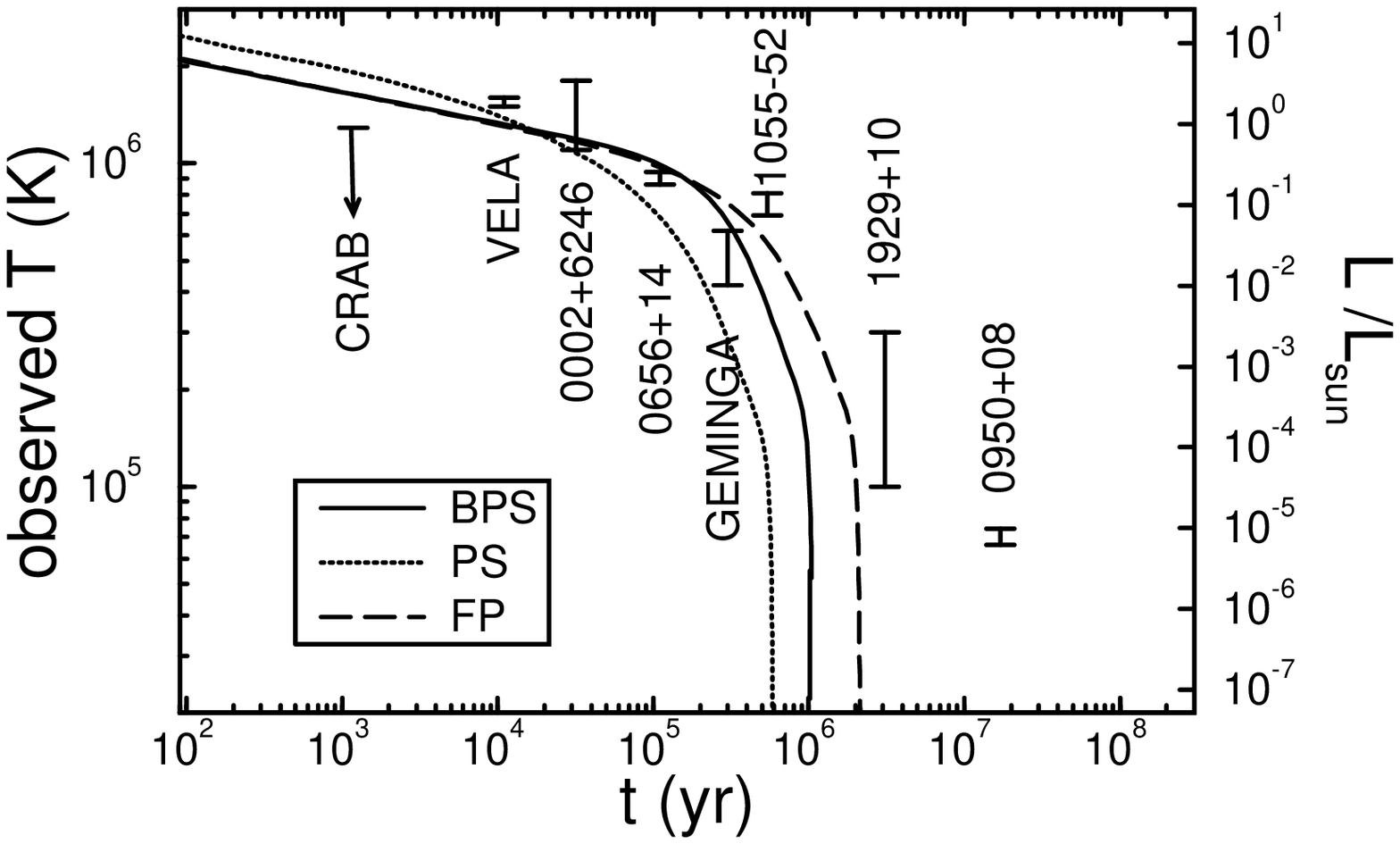}
\caption{
Thermal evolution curves for three different equations of state: FP
(\cite{FP}), PS (\cite{PS}) and BPS (\cite{BPS}).  These simulations
do not include internal heating.  Shown are surface temperatures and
luminosities measured by a distant observer. The observational temperature
determinations and the conversion to luminosity assume a stellar mass 
of 1.4 $M_{\odot}$ and radius $R = 10$ km.}
\end{figure}

\newpage

\begin{table}
\caption{${\rm Neutron \ Star \ Parameters^{a}}$}
\begin{tabular}{ll} \\ \hline\hline
$M$ (${\rm M_{\odot}}$) & 1.4 \\
$R$ (km) & 10.9 \\
$e^{-\Phi}$ & 1.27 \\
$I_s$ (g \ ${\rm cm^2}$) & $7.3 \times 10^{43}$ \\
$a$ (${\rm ergs \ K^{-2}}$) & $3.3 \times 10^{29}$ \\ \hline
\multicolumn{2}{l}{${\rm ^{a}}$Parameters based on the equation of 
state}\\
\multicolumn{2}{l}{of Friedman \& Pandharipande (1981).}
\label{bbbb}
\end{tabular}
\end{table}
	
\begin{deluxetable}{lcccll}
\tablecaption{Pinning Constraints for Old Stars} \tablewidth{40pc} 
\tablehead{ \colhead{Pulsar} & \colhead{$\log t_{\rm age}$} & 
\colhead{$\log T_{s,\infty}$} & \colhead{$\vert\dot\Omega_{0,\infty}\vert$} & 
\colhead{$\Delta J_{s}$} & \colhead{$\overline\omega$} \nl \colhead{} 
& \colhead{(yr)} & \colhead{(K)} & \colhead{(${\rm rad \ s^{-2}}$)} & 
\colhead{(ergs \ s)} & \colhead{(${\rm rad \ s^{-1}}$)}} 
\startdata 
1929+10 & 6.49 & 5.00 - 5.48 & 1.4 $\times 10^{-13}$ & 
9.8 $\times 10^{41}$ - 8.1 $\times 10^{43}$ & 0.01 - 1.1
 \tablenotemark{a}\nl 
0950+08 & 7.23 & 4.82 - 4.87 & 2.3 $\times 10^{-14}$ & 1.1 $\times 10^{42}$ 
- 1.8 $\times 10^{42}$ & 0.01 - 0.02 \nl 
1055-52 & 5.73 & 5.84 - 5.91 & 9.4 $\times 10^{-13}$ & 3.3 $\times 
10^{44}$ - 6.3 $\times 10^{44}$ & 4.6 - 8.6 \nl
\enddata 
\tablenotetext{a}{Uncertainties in the values of $\Delta J_{s}$ and 
$\overline\omega$ arise from uncertainties in the surface 
temperature (see \S 5).}
\label{dddd}
\end{deluxetable}

\begin{deluxetable}{lcclll}
\tablecaption{Pinning Constraints for Young Stars} \tablewidth{40pc} 
\tablehead{ \colhead{Pulsar} & \colhead{$\log t_{\rm age}$} & 
\colhead{$\log T_{s,\infty}$} & 
\colhead{$\omega_{0,{\rm max}}^{2}\Delta I_s$} & 
\colhead{$\overline\omega$\tablenotemark{a}} & \colhead{References} 
\nl \colhead{} & 
\colhead{(yr)} & \colhead{(K)} & \colhead{(ergs)} & 
\colhead{(${\rm rad \ s^{-1}}$)} & \colhead{}}
\startdata 
0531+21 & 3.10 & $<$ 6.19 & $ < 5.2\times 10^{46}$ & $<$ 35 & 
Becker \& Aschenback (1995) \nl 
0833-45 & 4.05 & 6.20 & $ < 5.4\times 10^{46}$ & $<$ 36 & 
\"Ogelman, Finley \& Zimmermann (1993)\tablenotemark{b} \nl 
0002+6246 & 4.50 & 6.26 & $ < 8.9\times 10^{46}$ & $<$ 46 &
Hailey \& Craig (1995)\tablenotemark{b}\nl 
0656+14 & 5.04 & 5.97 & $ < 7.7\times 10^{45}$ & $<$ 14 &
Finley, \"Ogelman \& Kizolo\u glu (1992)\nl 
0630+178 & 5.48 & 5.80 & $< 1.8 \times10^{45}$ & $<$ 6.6 &
Halpern \& Ruderman (1993)\nl 
1055-52 & 5.73 & 5.91 & $< 3.6 \times 10^{45}$ & $<$ 9.3 &
\"Ogelman \& Finley (1993)\nl 
\enddata 
\tablenotetext{a}{The upper limits on $\overline\omega$ assume 
$\overline\omega<\omega_{0,{\rm max}}$, as expected for most 
models of pinning.}
\tablenotetext{b}{The blackbody fits yield stellar radii of 2-4 km.}
\label{cccc}
\end{deluxetable}

\end{document}